\documentclass[a4paper]{article}

\usepackage{INTERSPEECH2021}
\usepackage{amsmath,graphicx}
\usepackage{amsthm,amsmath,amssymb}
\usepackage{mathrsfs}
\usepackage{booktabs}
\usepackage{threeparttable}
\usepackage{multirow,multicol}
\usepackage{makecell}
\usepackage[utf8]{inputenc}
\DeclareUnicodeCharacter{0002}{}
\title{Improving Accent Identification and Accented Speech Recognition Under a Framework of Self-supervised Learning}
\name{Keqi Deng$^{1,2,*}$, Songjun Cao$^{1,*}$, Long Ma$^1$ \thanks{ $^*$ Equal contribution.}}
\address{
  $^1$Tencent Cloud Xiaowei, Beijing, China\\
  $^2$University of Chinese Academy of Sciences, China}
\email{dengkeqi20@mails.ucas.ac.cn, \{songjuncao, malonema\}@tencent.com}

\begin{document}

\maketitle
\begin{abstract}
Recently, self-supervised pre-training has gained success in
automatic speech recognition (ASR).
However, considering the difference between speech accents in real scenarios, how to identify accents and use accent features to improve ASR is still challenging.
In this paper, we employ the self-supervised pre-training method for both accent identification and accented speech recognition tasks. 
For the former task, a standard deviation constraint loss (SDC-loss) based end-to-end (E2E) architecture is proposed to identify accents under the same language.
As for accented speech recognition task, we design an accent-dependent ASR system, which can utilize additional accent input features. Furthermore, we propose a frame-level accent feature, which is extracted based on the proposed accent identification model and can be dynamically adjusted.
We pre-train our models using 960 hours unlabeled LibriSpeech dataset and fine-tune them on AESRC2020 speech dataset.
The experimental results show that our proposed accent-dependent ASR system is significantly ahead of the AESRC2020 baseline and achieves $6.5\%$ relative word error rate (WER) reduction compared with our accent-independent ASR system.
\end{abstract}
\noindent\textbf{Index Terms}: self-supervised pre-training, accent identification, accented speech recognition

\section{Introduction}
In real scenarios, accent is one of the main and common sources of speech variability, which poses a huge challenge to automatic speech recognition (ASR). People coming from different countries or regions have their own distinctive accents and pronunciations. The differences between accents are mainly reflected in three aspects: stress, tone and length, which are challenging for ASR modeling \cite{shi2021accented}. Although different accents may share some similarities, there are obvious differences at the phonological level. As a result, the ASR system that trained on several kinds of accented speech simultaneously may fail to generalize well for each individual accent.
Therefore, it is beneficial to leverage the accent feature for the ASR model when recognizing accented speech.

However, since the accent category of the speech is not always provided in real scenarios, we still need an accent identification model to provide accent-related information and features for ASR systems. Compared with the individual-level identification task like speaker identification, accent identification throws a more challenging issue in extracting compact group-level features. Without a good discriminative accent feature space, over-fitting phenomenon always happens to accent identification. In addition, the phonological differences caused by accents are inconsistent in different words. Therefore, dynamically adjusting the accent-related information for the ASR system according to this inconsistency is still a valuable challenge. Furthermore, for the accent identification and accented speech recognition tasks, labeled data is much harder to get than unlabeled data. So it is meaningful to explore the self-supervised pre-training methods \cite{9054224, 9054438, peters-etal-2018-deep, Misra_2020_CVPR} to 
alleviate this problem in real scenarios.

In this paper, we propose a novel architecture for accent identification. Different from x-vector \cite{8461375,Snyder2017}, we directly identify the accent based on each frame rather than sentence-level vector, and then calculate the mean value of all frame-level outputs as the model's final prediction. 
In addition, we propose the standard deviation constraint loss (SDC-loss), which is based on the standard deviation of the frame-level outputs. The SDC-loss requires the predictions of each frame to be consistent, thereby curbing overfitting.
As for accented speech recognition, we design an accent-dependent ASR system, which can utilize additional accent input features. Furthermore, for situations where the ground truth of accent category is not provided, we propose a frame-level accent feature, which is extracted based on the proposed accent identification model and can be dynamically adjusted according to frame-level information.
We pre-train our models using 960 hours unlabeled LibriSpeech dataset following the wav2vec 2.0 \cite{NEURIPS2020_92d1e1eb} and fine-tune them on AESRC2020 \cite{shi2021accented} speech dataset. The experimental results show that our proposed accent-dependent ASR system can outperform all the previous baseline after fine-tuning.

The rest of this paper is organized as follows. In Section~\ref{sec:Relate}, we introduce the related works.
In Section~\ref{sec:proposed}, we describe the proposed architectures. The experiments and conclusions are presented in Sections~\ref{sec:experiments} and \ref{sec:con}, respectively.\par
\section{Related works}
\label{sec:Relate}
To distinguish different English accents, Teixeira et al. \cite{607975} propose contextual HMM units and Deshpande et al. \cite{1544415} employ format frequency features into GMM.
More recently, Shi et al. \cite{shi2021accented} propose an end-to-end (E2E) architecture for accent identification, but the phenomenon of over-fitting is still hard to avoid. 
Transfer learning and multi-task \cite{2020meng, Viglino2019} like using ASR downstream task to initialize the accent identification model is effective \cite{shi2021accented}. But this puts higher requirements on labeled data and self-supervised learning is more meaningful in real scenarios.

In general, accented ASR related research mainly falls into two ways: "multi-model" and "single-model" \cite{cao2021improving}. In multi-model approaches, an individual acoustic model is trained for each dialect accent when each accent's data is enough \cite{huang2014multi-accent, DBLP:conf/interspeech/ChenYLLL15}.
In single-model approaches, a single acoustic model is trained to deal with all dialect accents \cite{Jain2018, 7953071, 8683705}. And Li et al. \cite{8461886} incorporate accent information into a single ASR model by a 1-hot representation. But the representation they proposed is sentence-level and cannot be used in scenarios that do not provide true accent categories.
In this paper, we propose a frame-level accent feature and 
design a single accent-dependent ASR system to recognize all accented speech.

In machine learning, self-supervised learning has gained success in many fields \cite{DevlinCLT19, chen2020simple}.
Wav2vec \cite{Schneider2019} learns the representations of raw audio by solving a self-supervised context-prediction task.
In addition, Baevski et al. propose vq-wec2vec \cite{DBLP:conf/iclr/BaevskiSA20} that learns discrete speech representations through a context prediction task instead of reconstructing the input. Furthermore, Baevski et al. \cite{NEURIPS2020_92d1e1eb} propose wav2vec 2.0, which learns the discrete speech units and contextual representations end-to-end. In this paper, we employ the wav2vec 2.0 pre-training method.
\section{Proposed Method}
\label{sec:proposed}
Under the framework of wav2vec 2.0 \cite{NEURIPS2020_92d1e1eb}, our main focus is on fine-tuning stage. We first propose a SDC-loss based accent identification architecture. For accented speech recognition, we mainly explore how to utilize additional accent input features to improve accent-dependent ASR system. Both architectures contain a CNN based feature encoder, a Transformer based context network and a fully connected layer.

\subsection{Accent identification}
\begin{figure}[h]
    \centering
    \includegraphics[width=70mm]{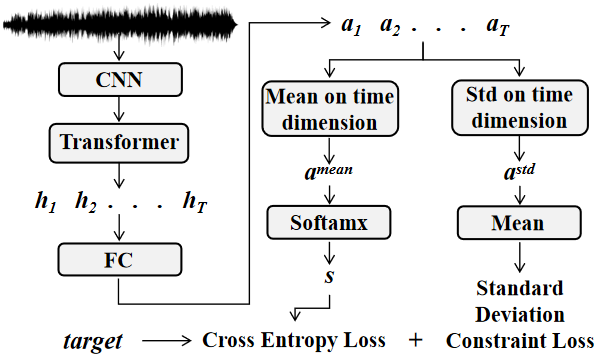}
    \caption{Illustration of the proposed accent identification architecture. }
    \label{fig:iden}
\end{figure}

The proposed accent identification model is shown in Fig.~\ref{fig:iden},
where ${\rm CNN}$ and ${\rm Transformer}$ denote CNN based feature encoder and Transformer based context network, respectively. $\mathbf{h}_{t}$ is the output of the Transformer \cite{Vaswani2017} at $t$ step. 
${\rm FC}$ is the fully connected layer and converts the  $\mathbf{h}_{t}$ to $\mathbf{a}_{t}$ whose dimension is equal to the number of accent categories. ${\rm Mean}$ and ${\rm Std}$ denote calculating the mean and standard deviation.

The proposed accent identification model directly predicts the accent category for each frame instead of sentence-level vector, and the final prediction is got by averaging.
Suppose $\mathbf{a}=(\mathbf{a}_{1} \cdot\cdot\cdot \mathbf{a}_{T})$ is the output of FC and $C$ denotes the number of total accent categories. We first calculate the mean and standard deviation of $\mathbf{a}$ and denote them as $\mathbf{a}^{mean}$ and $\mathbf{a}^{std}$, respectively.
Then the SDC-loss is defined as:
\begin{equation}
     \mathcal{L}_{SDC}=\frac{1}{C}\sum_{j=1}^{C}{a^{std}_{j}}
\end{equation}
We set $\mathbf{a}^{mean}$ as the final prediction, and compare it with the target after softmax to obtain the cross entropy loss (CE-loss). 
\begin{equation}
     \mathcal{L}_{CE}={\rm Cross\text{-}entropy}(\mathbf{a}^{mean}, \mathbf{g}^{true})
\end{equation}
where $\mathbf{g}^{true}$ represents the target and $\mathcal{L}_{CE}$ denotes the CE-loss. The final loss function is obtained by adding the SDC-loss and CE-loss together.
\begin{equation}
     \mathcal{L}_{final}=\mathcal{L}_{CE}+\mathcal{L}_{SDC},\label{final}
\end{equation}
where $\mathcal{L}_{final}$ denotes the final loss function.
The $\mathcal{L}_{final}$ requires not only that the final prediction matches the target, but also that the difference between the predictions of each frame to be small. This can not only suppress over-fitting, but also has further significance for extracting frame-level accent features for ASR system, which will be explained in next section.

\subsection{Accented speech recognition}
\begin{figure}[h]
    \centering
    \includegraphics[width=65mm]{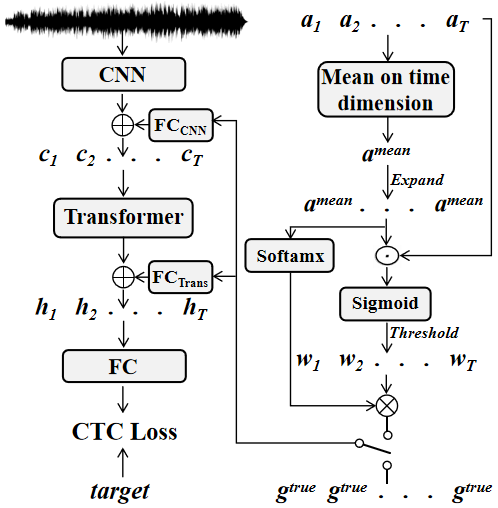}
    \caption{Illustration of the proposed accented speech recognition architecture. }
    \label{fig:asr}
\end{figure}

The proposed accented speech recognition model is based on CTC and
shown in Fig.~\ref{fig:asr}, where $\mathbf{a_{i}}$ is the output of accent identification model as shown in Fig.~\ref{fig:iden}. ${\rm Expand}$ means broadcasting the data to expand the size same as the dot multiplied object. \textcircled{·}, \textcircled{×} and \textcircled{+} denote dot product, scalar multiplication and point-wise plus respectively. 
${\rm FC_{CNN}}$ and ${\rm FC_{Trans}}$ are the fully connected layer for ${\rm CNN}$ and ${\rm Transformer}$ respectively. $\mathbf{g}^{true}$ is a one-hot vector, which corresponds to the ground truth of the utterance's accent category, and we expand its size as frame level.

But in the real scenarios, the true accent category is not always provided. Therefore, we need to use the accent identification model to generate accent-related information to improve the accent-dependent ASR system. Next, we will separately describe the accented speech recognition in these two conditions.
\subsubsection{Accented speech recognition with true accent category}
Supposed the outputs of ${\rm CNN}$ and ${\rm Transformer}$ are $\mathbf{\hat{c}}_{i}$ and $\mathbf{\hat{h}}_{i}$,
where $i \in (1,T)$. When the true accent category is provided, the $\mathbf{g}^{true}$ is linearly transformed by the ${\rm FC_{CNN}}$ and added to $\mathbf{\hat{c}}_{i}$. Similarly, we also add it to the $\mathbf{\hat{h_{i}}}$ through the ${\rm FC_{Trans}}$.
\begin{eqnarray}
     \mathbf{c}_{i}&=&\mathbf{\hat{c}}_{i}+{\rm FC_{CNN}}(\mathbf{g}^{true}),\\
     \mathbf{h}_{i}&=&\mathbf{\hat{h}}_{i}+{\rm FC_{Trans}}(\mathbf{g}^{true}),
\end{eqnarray}
This effectively enables the accent-dependent ASR system to learn accent-related biases.
\subsubsection{Accented speech recognition without true accent category}
In this part, we describe our proposed accent-dependent ASR system when the true accent category is not provided. In traditional method, 
the accent-related features obtained from the ground truth of accent category are sentence-level \cite{8683705, 8461886}, which have no specific requirement for frame-level vectors before pooling.
However, since the accent identification model we proposed is based on frame-level vector and the SDC-loss requires the prediction of each frame match the final prediction, we can get frame-level information. We believe that in accented speech recognition, it is not necessary to add accent-related bias for every frame. Instead, different weights should be given according to the importance of each frame for accent identification.  In this paper, we propose frame-level accent feature, which utilizes the frame-level information to adjust the weight of the accent bias for each frame. The details are as follows:
\begin{equation}
     {w_{i}}={\rm Sigmoid}(\mathbf{a}_{i} \cdot \mathbf{a}^{mean}), \label{w}
\end{equation}
where $\cdot$ is the dot product and $w_{i}$ is the weight of the $i$-th accent feature. We also set a threshold $k$ for $w_{i}$:
\begin{equation}
w_{i}=
\begin{cases}
w_{i},& \text{ $ w_{i} \textgreater k$ } \\
0,& \text{ $ w_{i} \textless k$} \label{threshold}
\end{cases}
\end{equation}
Finally, the accented-related bias is added to the output of the CNN and Transformer:
\begin{eqnarray}
     \mathbf{c}_{i}&=&\mathbf{\hat{c}_{i}}+{\rm FC_{CNN}}(w_{i} \cdot {\rm Softmax}(\mathbf{a}^{mean})),\\
     \mathbf{h}_{i}&=&\mathbf{\hat{h}_{i}}+{\rm FC_{Trans}}(w_{i} \cdot {\rm Softmax}(\mathbf{a}^{mean})),\label{hhat}
\end{eqnarray}
\section{Experiments}
\label{sec:experiments}
\subsection{Corpus}
We pre-train both the accent identification model and the accented speech recognition model on the Librispeech corpus \cite{7178964} without transcriptions containing 960 hours of audio (LS-960). And we fine-tune the models on the AESRC2020 speech corpus \cite{shi2021accented}. The AESRC2020 speech corpus contains 160 hours speech data and includes 8 accents: English, America, China, Japan, Russia, India, Portugal, Korea. The training set contains around 120000 utterances and the testing set we used consists of around 12000 utterances.
\subsection{Model descriptions}
We use the Fairseq toolkit \cite{ott2019fairseq} to build the models. For the acoustic input, we employ the waveform following wav2vec 2.0 \cite{NEURIPS2020_92d1e1eb}. For the text output, we used a 28 vocabulary set, including 26 English characters and 2 punctuations.

CNN based feature encoder consists of seven convolution layers (i.e., 512 channels with kernel size 10,3,3,3,3,2 and 2 and strides 5,2,2,2,2,2 and 2).
Transformer based context network contains 12 Transformer layers with 768 model dimensions, 3072 inner dimensions (FFN) and eight attention heads.
The dimension of ${\rm FC}$ in accent identification model is 8.
In accented ASR system, the dimensions of the ${\rm FC_{CNN}}$ and ${\rm FC_{Trans}}$ are 512 and 768, respectively. And the threshold $k$ in Eq.~\ref{threshold} is set to 0.4.

During pre-training, we follow by Fairseq recipe \cite{ott2019fairseq}. During fine-tuning, we use a batch size of 3.2M samples per GPU. For accent identification tasks, we fine-tune the model by the Adam  \cite{DBLP:journals/corr/KingmaB14} optimizer (learning rate 2e-5) with 1600 warm steps on 4 GPUs. The max update number is 3000 and for the first 2000 updates only the $FC$ is optimized,
after which the Transformer based context network is also updated. The CNN based feature encoder is fixed during fine-tuning. As for accented speech recognition task, we also employ the Adam optimizer (learning rate 2e-5) with 8000 warm steps and we train the model on 8 GPUs. The max update number is 40000 and all the networks can be updated except the CNN based feature encoder. In order to prevent overfitting, we used dropout \cite{Nitish2014dropout} (dropout rate=0.1) after each sub-layer of the Transformer based context network and the output of the CNN based feature encoder. We also employ the layer dropout method (dropout rate=0.1). During decoding, the beam size is fixed as 500 and the word insertion penalty \cite{7015055} is -0.52.
\subsection{Accent identification results}

First, we extract the i-vector \cite{5545402} and x-vector \cite{8461375} using ASV-Subtools \cite{mta}, based on which we train a logistic regression model for accent identification.
Second, based on the pre-trained model, a mean + std pooling layer is applied after ${\rm Transformer}$ to pool the $\mathbf{h}_{t}$, which is the same as the method \cite{shi2021accented}.
We denote this model as ${\rm \textbf{AI}_{1}}$ and train it end-to-end. 
At last, we denote the proposed SDC-loss based accent identification model as ${\rm \textbf{AI}_{2}}$. As for ablation study, we do not introduce the SDC-loss of ${\rm \textbf{AI}_{2}}$ and mark it as ${\rm \textbf{AI}_{3}}$. The accent identification results are as shown in Table~\ref{tab:PC}, where B, A, C, J, R, I, P, K represent the accent of British, America, China, Japan, Russia, India, Portugal, Korea. AESRC-12L and AESRC-3L denote the results of 12-layer and 3-layer Transformer in AESRC2020 \cite{shi2021accented}. And the "+ASR init" employs the ASR initialization method \cite{shi2021accented} based on the "AESRC-12L". 
\begin{table}[h]
  \caption{The accuracy (\%) of the accent identification on AESRC2020.}
  \setlength\tabcolsep{1.7pt}
  \label{tab:PC}
  \centering
  \begin{tabular}{c c c c c c c c c c}
    \toprule
    Method&
    B&
    A&
    C&
    J&
    R&
    I&
    P&
    K&
    All\\
    \hline
    AESRC-12L & 85.0 &21.2  & 38.2 &42.7 & 49.6& 66.1&51.8 &26.0&47.8 \\
    AESRC-3L & 70.0 &45.7  & 56.2 &48.5 & 30.0& 83.5&57.2 &45.0&54.1 \\
    +ASR init & 93.9 &60.2  & 67.0 &73.2 & 75.7& 97.0&85.5 &55.6&76.1 \\
    \hline
    x-vector \cite{8461375} &  66.0 & 46.4 & 62.2 & 53.5 & 59.9 & 77.4 & 54.7 & 56.1 & 59.1 \\
    i-vector \cite{5545402} & 72.1 & 48.8 & 70.9 & 45.9 & 64.3 & 87.4 & 58.0 & 58.1 & 62.6\\
    ${\rm {AI}_{1}}$ & 67.7 & 87.2 & 61.5 & 94.8 &51.4 &63.0 &82.0 &74.7 & 72.7\\
    ${\rm {AI}_{2}}$ & 83.2 &  89.6& 54.0 &95.8  &52.6 &65.2 &82.7 &70.2 & \textbf{73.9}\\
    ${\rm {AI}_{3}}$ &76.3  & 87.2 & 61.5 &94.8  & 51.41& 63.0&82.1&74.7 & 72.9\\
    \bottomrule
  \end{tabular}
\end{table}

The results in Table~\ref{tab:PC} show that, based on the self-supervised pre-trained model, end-to-end fine-tuning outperforms the vanilla x-vector and i-vector methods. In addition, the ${\rm \textbf{AI}_{1}}$ are sentence-level accent identification model, while the ${\rm \textbf{AI}_{2}}$ and ${\rm \textbf{AI}_{3}}$ identify the accent based on frame-level vector. From the results we can see that the proposed method ${\rm \textbf{AI}_{2}}$ outperforms the sentence-level model ${\rm \textbf{AI}_{1}}$. Furthermore, through the ablation study, the proposed SDC-loss that encourages the prediction of each frame to be consistent does improve the model's identification performance.
It puts forward higher requirements on the model's prediction, thereby curbing overfitting. At last, although AESRC-12L has more parameters than AESRC-3L, due to over-fitting, its results are worse. Our proposed ${\rm \textbf{AI}_{2}}$ also contains 12-layer Transformer, but due to the self-supervised pre-training method and the proposed SDC-loss,
${\rm \textbf{AI}_{2}}$ greatly outperforms AESRC-3L and AESRC-12L.
After employing the ASR initialization method \cite{shi2021accented}, AESRC-12L's performance has been greatly improved. But we think that this method requires more supervised labeled information and the self-supervised pre-training method is more meaningful in real scenarios.
And under the framework of self-supervised learning, we get close results.

\subsection{Accented speech recognition results}
As for accented speech recognition, based on the pre-trained model, we
apply a ${\rm FC}$ and fine-tune it based on CTC loss. We
mainly investigate utilizing the additional accent input features obtained from ground truth or the proposed accent identification model for the accent-dependent ASR system.
\subsubsection{With true accent category}
With the true accent category, we do not need an accent identification model.
First, we train an accent-independent ASR model, which ignores the accent category and is denoted as ${\rm \textbf{AR}_{0}}$. And we denote the proposed architecture as ${\rm \textbf{AR}_{1}}$, which chooses $\mathbf{g}^{true}$ as accent vector and is illustrated in Fig.~\ref{fig:asr}. As for ablation study, 1. adding the accent-related bias only after the Transformer based context network (${\rm \textbf{AR}_{2}}$); 2. concatenating the output of the ${\rm FC_{Trans}}$ with the $\mathbf{\hat{h_{i}}}$ as $\mathbf{h}_{i}$ in Eq.~\ref{hhat} (${\rm \textbf{AR}_{3}}$); 3. taking the ${\rm \textbf{AR}_{0}}$ as starting point and only updating accent-dependent output layers (eight in total and one for each accent) for extra 3500 update numbers (${\rm \textbf{AR}_{4}}$).
\begin{table}[h]
  \caption{The WER (\%) of the accented speech recognition on AESRC2020 with true accent category provided.}
  \setlength\tabcolsep{1.8pt}
  \label{asr}
  \centering
  \begin{tabular}{c c c c c c c c c c}
    \toprule
    Method&
    B&
    A&
    C&
    J&
    R&
    I&
    P&
    K&
    All\\
    \hline
    ${\rm {AR}_{0}}$ &7.50  & 6.29 &10.64  & 8.21 &7.46 & 7.86& 6.22&5.00 & 7.37\\
    ${\rm {AR}_{1}}$ &7.44  &5.99  &9.67  &7.48  &7.13 &7.07 &5.86 & 4.72& \textbf{6.89}\\
    ${\rm {AR}_{2}}$ & 7.15 & 5.91 &10.81  &7.74  &7.11 &7.71 & 5.87&4.68 & 7.09\\
    ${\rm {AR}_{3}}$ & 7.48 &6.05  &10.69  &7.96  &7.12 &7.64 &5.91 &4.78 & 7.17\\
    ${\rm {AR}_{4}}$ & 7.21 &6.18  &10.64  &7.91  &7.14 &7.58 &6.04 &4.95 & 7.18\\
    \bottomrule
  \end{tabular}
\end{table}

The results in Table~\ref{tab:PC} show that 
the proposed ${\rm \textbf{AR}_{1}}$ achieves $6.4\%$ relative word error rate (WER) reduction compared with the ${\rm \textbf{AR}_{0}}$, which proves that the accent-related bias does improve the ASR system's performance on accented speech. Through ablation study, we found that 1. adding the accent-related bias after both CNN and Transformer achieves the best; 2. adding the accent-related bias to the $\mathbf{\hat{h_{i}}}$ outperforms concatenating them.
3. the improvement of employing accent-dependent output layers is limited.
\subsubsection{Without true accent category}
In real scenarios, the true accent category is not always provided.
In this part, we utilize the accent identification model to produce the accent-related feature for the ASR. We denote our proposed architecture shown in Fig.~\ref{fig:asr} as ${\rm \textbf{AR}_{5}}$. As for ablation, 
1. we do not introduce the threshold $k$ on the basis of ${\rm \textbf{AR}_{5}}$ and denote it as ${\rm \textbf{AR}_{6}}$; 2. we do not introduce the weight of accent-related bias $w_{i}$ in Eq.~\ref{w} and denote it as ${\rm \textbf{AR}_{7}}$. In this ways, ${\rm \textbf{AR}_{7}}$
can be regarded as all $w_{i}$ equals to 1 and thus
can not utilize the frame-level information. 
\begin{table}[h]
  \caption{The WER (\%) of the accented speech recognition on AESRC2020 without true accent category provided.}
  \setlength\tabcolsep{2.5pt}
  \label{asr2}
  \centering
  \begin{tabular}{c c c c c c c c c c}
    \toprule
    Method&
    B&
    A&
    C&
    J&
    R&
    I&
    P&
    K&
    All\\
    \hline
    ${\rm {AR}_{1}}$ &7.44  &5.99  &9.67  &7.48  &7.13 &7.07 &5.86 & 4.72& 6.89\\
    \hline
    ${\rm {AR}_{5}}$ & 6.95 &5.85  &  10.52& 7.84 &6.92 &7.80 &5.82 &4.72 & \textbf{7.02}\\
    ${\rm {AR}_{6}}$ & 7.22 & 6.01 &  10.44& 7.92 & 6.99&7.66 &6.01 &4.69 & 7.09\\
    ${\rm {AR}_{7}}$ &7.18& 5.98 & 10.46 & 8.12 &7.19  & 7.67& 6.14& 4.84& 7.17\\
    \bottomrule
  \end{tabular}
\end{table}

From the results in Table~\ref{asr2}, we can see that with the proposed frame-level accent features, the ASR system can get close to the result that is provided the ground truth of accent categories. 
In addition, through the ablation study, we can prove that providing frame-level information does help the accent-related bias better improve the performance of ASR on accented speech.
And using the weight $w_{i}$ to represent the importance of each frame for accent identification and dynamically adjusting the accent-related bias outperform adding the same accent-related bias for each frame. 
\subsubsection{Comparison with other methods}
In this part, we compare our method with the results in AESRC \cite{shi2021accented}. The results are as follows, where the "AESRC" represents the result of only using the labeled data from AESRC2020 speech corpus with RNN language model (LM) and the "+LS-960" uses additional labeled data from the LS-960 dataset.
\begin{table}[h]
  \caption{WER (\%) comparison of different methods on AESRC.}
  \setlength\tabcolsep{1.7pt}
  \label{other}
  \centering
  \begin{tabular}{c c c c c c c c c c}
    \toprule
    Method&
    B&
    A&
    C&
    J&
    R&
    I&
    P&
    K&
    All\\
    \hline
    AESRC & 10.06 &9.96  & 11.77 &6.79 & 5.26& 10.05&7.45 &7.69&8.63 \\
    +LS-960 & 7.64 &7.42  & 9.87 &5.71 & 4.60&7.85&5.90 &6.40&6.92 \\
    \hline
    ${\rm {AR}_{1}}$ &7.44  &5.99  &9.67  &7.48  &7.13 &7.07 &5.86 & 4.72& \textbf{6.89}\\
    +$4$-gram &4.81  &4.06  &7.09  &4.51  &4.44 &4.22 &3.73 & 2.55& \textbf{4.42}\\
    \hline
    ${\rm {AR}_{5}}$ & 6.95 &5.85  &  10.52& 7.84 &6.92 &7.80 &5.82 &4.72 & \textbf{7.02}\\
    +$4$-gram & 4.68 &4.20  &  7.60& 4.72 &4.33 &4.47 &3.71 &2.57 & \textbf{4.52}\\
    \bottomrule
  \end{tabular}
\end{table}

It can be seen from the results that self-supervised pre-training is effective for accented speech recognition. Except for AESRC2020 dataset, we only use the unlabeled data from LS-960 dataset additionally. Without using LM, we can still surpass the results of "AESRC". And we train a 4-gram LM \cite{Goodman2001} on the same data as RNN LM: transcription of LS-960 and AESRC2020 datasets.
After employing the 4-gram LM (weight: 1.74),
we achieve $36.1\%$ relative WER reduction.  
This also shows that LM is of great help to the CTC-based ASR system, as CTC model has conditional independence assumption.

\section{Conclusion}
\label{sec:con}
In this paper, we explore the self-supervised pre-training methods to solve the accent identification and accented speech recognition tasks. 
Based on the pre-trained model following wav2vec 2.0,
we propose a SDC-loss based E2E architecture 
to identify accents under the same language. As for accented speech recognition, we design an accent-dependent ASR system, which can utilize additional accent input features extracted from the ground truth of accent category or
the proposed accent identification model.
Furthermore, 
we propose a frame-level accent feature, which is extracted based on the proposed accent identification model and can leverage the frame-level information.
We pre-train the networks using 960 hours unlabeled LibriSpeech dataset and fine-tune them on AESRC2020 speech dataset. The experimental results show that our proposed accent-dependent ASR system is significantly ahead of the AESRC2020 baseline and achieves $6.5\%$ relative WER reduction compared with our accent-independent ASR system.
\bibliographystyle{IEEEtran}

\bibliography{mybib}

\begin{thebibliography}{10}
\providecommand{\url}[1]{#1}
\csname url@samestyle\endcsname
\providecommand{\newblock}{\relax}
\providecommand{\bibinfo}[2]{#2}
\providecommand{\BIBentrySTDinterwordspacing}{\spaceskip=0pt\relax}
\providecommand{\BIBentryALTinterwordstretchfactor}{4}
\providecommand{\BIBentryALTinterwordspacing}{\spaceskip=\fontdimen2\font plus
\BIBentryALTinterwordstretchfactor\fontdimen3\font minus
  \fontdimen4\font\relax}
\providecommand{\BIBforeignlanguage}[2]{{%
\expandafter\ifx\csname l@#1\endcsname\relax
\typeout{** WARNING: IEEEtran.bst: No hyphenation pattern has been}%
\typeout{** loaded for the language `#1'. Using the pattern for}%
\typeout{** the default language instead.}%
\else
\language=\csname l@#1\endcsname
\fi
#2}}
\providecommand{\BIBdecl}{\relax}
\BIBdecl

\bibitem{shi2021accented}
X.~Shi, F.~Yu, Y.~Lu, Y.~Liang, Q.~Feng, D.~Wang, Y.~Qian, and L.~Xie, ``The
  accented english speech recognition challenge 2020: Open datasets, tracks,
  baselines, results and methods,'' 2021.

\bibitem{9054224}
A.~{Baevski} and A.~{Mohamed}, ``Effectiveness of self-supervised pre-training
  for asr,'' in \emph{ICASSP 2020 - 2020 IEEE International Conference on
  Acoustics, Speech and Signal Processing (ICASSP)}, 2020, pp. 7694--7698.

\bibitem{9054438}
Y.~A. {Chung} and J.~{Glass}, ``Generative pre-training for speech with
  autoregressive predictive coding,'' in \emph{ICASSP 2020 - 2020 IEEE
  International Conference on Acoustics, Speech and Signal Processing
  (ICASSP)}, 2020, pp. 3497--3501.

\bibitem{peters-etal-2018-deep}
M.~Peters, M.~Neumann, M.~Iyyer, M.~Gardner, C.~Clark, K.~Lee, and
  L.~Zettlemoyer, ``Deep contextualized word representations,'' in
  \emph{Proceedings of the 2018 Conference of the North {A}merican Chapter of
  the Association for Computational Linguistics: Human Language Technologies,
  Volume 1 (Long Papers)}, Jun. 2018, pp. 2227--2237.

\bibitem{Misra_2020_CVPR}
I.~Misra and L.~v.~d. Maaten, ``Self-supervised learning of pretext-invariant
  representations,'' in \emph{Proceedings of the IEEE/CVF Conference on
  Computer Vision and Pattern Recognition (CVPR)}, June 2020.

\bibitem{8461375}
D.~{Snyder}, D.~{Garcia-Romero}, G.~{Sell}, D.~{Povey}, and S.~{Khudanpur},
  ``X-vectors: Robust dnn embeddings for speaker recognition,'' in \emph{2018
  IEEE International Conference on Acoustics, Speech and Signal Processing
  (ICASSP)}, 2018, pp. 5329--5333.

\bibitem{Snyder2017}
D.~Snyder, D.~Garcia-Romero, D.~Povey, and S.~Khudanpur, ``Deep neural network
  embeddings for text-independent speaker verification,'' in \emph{Proc.
  Interspeech 2017}, 2017, pp. 999--1003.

\bibitem{NEURIPS2020_92d1e1eb}
A.~Baevski, Y.~Zhou, A.~Mohamed, and M.~Auli, ``wav2vec 2.0: A framework for
  self-supervised learning of speech representations,'' in \emph{Advances in
  Neural Information Processing Systems}, H.~Larochelle, M.~Ranzato,
  R.~Hadsell, M.~F. Balcan, and H.~Lin, Eds., vol.~33.\hskip 1em plus 0.5em
  minus 0.4em\relax Curran Associates, Inc., 2020, pp. 12\,449--12\,460.

\bibitem{607975}
C.~{Teixeira}, I.~{Trancoso}, and A.~{Serralheiro}, ``Accent identification,''
  in \emph{Proceeding of Fourth International Conference on Spoken Language
  Processing. ICSLP '96}, vol.~3, 1996, pp. 1784--1787 vol.3.

\bibitem{1544415}
S.~{Deshpande}, S.~{Chikkerur}, and V.~{Govindaraju}, ``Accent classification
  in speech,'' in \emph{Fourth IEEE Workshop on Automatic Identification
  Advanced Technologies (AutoID'05)}, 2005, pp. 139--143.

\bibitem{2020meng}
Z.~{Meng}, H.~{Hu}, J.~{Li}, C.~{Liu}, Y.~{Huang}, Y.~{Gong}, and C.~{Lee},
  ``L-vector: Neural label embedding for domain adaptation,'' in \emph{ICASSP
  2020 - 2020 IEEE International Conference on Acoustics, Speech and Signal
  Processing (ICASSP)}, 2020, p. 7389–7393.

\bibitem{Viglino2019}
T.~Viglino, P.~Motlicek, and M.~Cernak, ``{End-to-End Accented Speech
  Recognition},'' in \emph{Proc. Interspeech 2019}, 2019, pp. 2140--2144.

\bibitem{cao2021improving}
S.~Cao, Y.~Zhang, X.~Feng, and M.~Long, ``Improving speech recognition accuracy
  of local poi using geographical models,'' in \emph{2021 IEEE Spoken Language
  Technology Workshop (SLT)}.\hskip 1em plus 0.5em minus 0.4em\relax IEEE,
  2021.

\bibitem{huang2014multi-accent}
Y.~Huang, D.~Yu, C.~Liu, and Y.~Gong, ``Multi-accent deep neural network
  acoustic model with accent-specific top layer using the kld-regularized model
  adaptation,'' in \emph{Interspeech 2014}, September 2014.

\bibitem{DBLP:conf/interspeech/ChenYLLL15}
M.~Chen, Z.~Yang, J.~Liang, Y.~Li, and W.~Liu, ``Improving deep neural networks
  based multi-accent mandarin speech recognition using i-vectors and
  accent-specific top layer,'' in \emph{{INTERSPEECH} 2015, 16th Annual
  Conference of the International Speech Communication Association, Dresden,
  Germany, September 6-10, 2015}.\hskip 1em plus 0.5em minus 0.4em\relax
  {ISCA}, 2015, pp. 3620--3624.

\bibitem{Jain2018}
A.~Jain, M.~Upreti, and P.~Jyothi, ``Improved accented speech recognition using
  accent embeddings and multi-task learning,'' in \emph{Proc. Interspeech
  2018}, 2018, pp. 2454--2458.

\bibitem{7953071}
K.~{Rao} and H.~{Sak}, ``Multi-accent speech recognition with hierarchical
  grapheme based models,'' in \emph{2017 IEEE International Conference on
  Acoustics, Speech and Signal Processing (ICASSP)}, 2017, pp. 4815--4819.

\bibitem{8683705}
S.~{Yoo}, I.~{Song}, and Y.~{Bengio}, ``A highly adaptive acoustic model for
  accurate multi-dialect speech recognition,'' in \emph{ICASSP 2019 - 2019 IEEE
  International Conference on Acoustics, Speech and Signal Processing
  (ICASSP)}, 2019, pp. 5716--5720.

\bibitem{8461886}
B.~{Li}, T.~N. {Sainath}, K.~C. {Sim}, M.~{Bacchiani}, E.~{Weinstein},
  P.~{Nguyen}, Z.~{Chen}, Y.~{Wu}, and K.~{Rao}, ``Multi-dialect speech
  recognition with a single sequence-to-sequence model,'' in \emph{2018 IEEE
  International Conference on Acoustics, Speech and Signal Processing
  (ICASSP)}, 2018, pp. 4749--4753.

\bibitem{DevlinCLT19}
J.~Devlin, M.~Chang, K.~Lee, and K.~Toutanova, ``{BERT:} pre-training of deep
  bidirectional transformers for language understanding,'' in \emph{Proceedings
  of the 2019 Conference of the North American Chapter of the Association for
  Computational Linguistics: Human Language Technologies, {NAACL-HLT} 2019,
  Minneapolis, MN, USA, June 2-7, 2019, Volume 1 (Long and Short
  Papers)}.\hskip 1em plus 0.5em minus 0.4em\relax Association for
  Computational Linguistics, 2019, pp. 4171--4186.

\bibitem{chen2020simple}
T.~Chen, S.~Kornblith, M.~Norouzi, and G.~Hinton, ``A simple framework for
  contrastive learning of visual representations,'' in \emph{International
  conference on machine learning}.\hskip 1em plus 0.5em minus 0.4em\relax PMLR,
  2020, pp. 1597--1607.

\bibitem{Schneider2019}
S.~Schneider, A.~Baevski, R.~Collobert, and M.~Auli, ``{wav2vec: Unsupervised
  Pre-Training for Speech Recognition},'' in \emph{Proc. Interspeech 2019},
  2019, pp. 3465--3469.

\bibitem{DBLP:conf/iclr/BaevskiSA20}
A.~Baevski, S.~Schneider, and M.~Auli, ``vq-wav2vec: Self-supervised learning
  of discrete speech representations,'' in \emph{8th International Conference
  on Learning Representations, {ICLR} 2020, Addis Ababa, Ethiopia, April 26-30,
  2020}, 2020.

\bibitem{Vaswani2017}
A.~Vaswani, N.~Shazeer, N.~Parmar, J.~Uszkoreit, L.~Jones, A.~N. Gomez,
  L.~Kaiser, and I.~Polosukhin, ``Attention is all you need,'' in
  \emph{Proceedings of the 31st International Conference on Neural Information
  Processing Systems}, ser. NIPS'17.\hskip 1em plus 0.5em minus 0.4em\relax
  USA: Curran Associates Inc., 2017, pp. 6000--6010.

\bibitem{7178964}
V.~{Panayotov}, G.~{Chen}, D.~{Povey}, and S.~{Khudanpur}, ``Librispeech: An
  asr corpus based on public domain audio books,'' in \emph{2015 IEEE
  International Conference on Acoustics, Speech and Signal Processing
  (ICASSP)}, 2015, pp. 5206--5210.

\bibitem{ott2019fairseq}
M.~Ott, S.~Edunov, A.~Baevski, A.~Fan, S.~Gross, N.~Ng, D.~Grangier, and
  M.~Auli, ``fairseq: A fast, extensible toolkit for sequence modeling,'' in
  \emph{Proceedings of NAACL-HLT 2019: Demonstrations}, 2019.

\bibitem{DBLP:journals/corr/KingmaB14}
D.~P. Kingma and J.~Ba, ``Adam: {A} method for stochastic optimization,'' in
  \emph{3rd International Conference on Learning Representations, {ICLR} 2015,
  San Diego, CA, USA, May 7-9, 2015, Conference Track Proceedings}, 2015.

\bibitem{Nitish2014dropout}
N.~Srivastava, G.~Hinton, A.~Krizhevsky, I.~Sutskever, and R.~Salakhutdinov,
  ``Dropout: A simple way to prevent neural networks from overfitting,''
  \emph{J. Mach. Learn. Res.}, vol.~15, no.~1, pp. 1929--1958, Jan. 2014.

\bibitem{7015055}
T.~{Valenta} and L.~{Šmídl}, ``On the impact of sentence length on
  recognition accuracy,'' in \emph{2014 12th International Conference on Signal
  Processing (ICSP)}, 2014, pp. 500--504.

\bibitem{5545402}
N.~{Dehak}, P.~J. {Kenny}, R.~{Dehak}, P.~{Dumouchel}, and P.~{Ouellet},
  ``Front-end factor analysis for speaker verification,'' \emph{IEEE
  Transactions on Audio, Speech, and Language Processing}, vol.~19, no.~4, pp.
  788--798, 2011.

\bibitem{mta}
M.~Zhao, J.~Zhou, Z.~Li, H.~Lu, and F.~Tong, ``Asv-subtools: An open source
  tools for speaker recognition,'' https://github.com/Snowdar/asv-subtools,
  2021, gitHub repository.

\bibitem{Goodman2001}
J.~T. Goodman, ``A bit of progress in language modeling,'' \emph{Computer
  Speech \& Language}, vol.~15, pp. 403--434, 2001.

\end{thebibliography}


\end{document}